\documentclass[aps,prb,twocolumn,showpacs,superscriptaddress,floatfix,reprint]{revtex4-2}

\usepackage{amsmath,amssymb}
\usepackage{graphicx}% Include figure files
\usepackage{dcolumn}% Align table columns on decimal point
\usepackage{bm}% bold math
\usepackage{comment}
\usepackage{xcolor}

\begin{document}

\preprint{APS/123-QED}

\title{Nondegenerate two-photon lasing in a single quantum dot}

\author{Samit Kumar Hazra}
\email{samit176121009@iitg.ac.in}
\affiliation{Department of Physics, Indian Institute of Technology Guwahati, Guwahati 781039, Assam, India}
\author{Lava Kumar Addepalli}
\email{d20034@students.iitmandi.ac.in}
\affiliation{Indian Institute of Technology Mandi, Mandi 175001, Himachal Pradesh, India}
\author{P. K. Pathak}
\email{ppathak@iitmandi.ac.in}
\affiliation{Indian Institute of Technology Mandi, Mandi 175001, Himachal Pradesh, India}
\author{Tarak Nath Dey}
\email{tarak.dey@iitg.ac.in}
\affiliation{Department of Physics, Indian Institute of Technology Guwahati, Guwahati 781039, Assam, India}

\date{\today}% It is always \today, today,

\begin{abstract}
We propose two-mode two-photon microlaser using a single semiconductor quantum dot grown inside a two-mode microcavity.  We explore both incoherent and coherent pumping at low temperatures to achieve suitable conditions for two-mode two-photon lasing. The two-mode two-photon stimulated emission is strongly suppressed but the single-photon stimulated emission is enhanced by exciton-phonon interactions.  In coherently pumped quantum dot one can achieve large two-mode two-photon lasing where single-photon lasing is almost absent. We also discuss generation of steady state two-mode entangled state using two-photon resonant pumping. 
\end{abstract}

\maketitle

%\tableofcontents

\section{\label{sec:level1}INTRODUCTION
%First-level heading:\protect\\ The line
%break was forced \lowercase{via} \textbackslash\textbackslash
}

The quantum theory of maser was established in the late 20th century \cite{QTL1,QTL2}, in which gain medium has been modelled as quantum emitters interacting with the quantized cavity field. The single atom maser was experimentally demonstrated using Rydberg atoms \cite{OAM}. Since then, following developments in quantum optics, single-atom lasers have been realized using several other quantum emitters coupled with different high-quality cavities, including trapped ions \cite{SIL}, trapped single atom \cite{SAQL}, superconducting qubits \cite{SAAL1,SAAL2}, and quantum dots \cite{QDL1,QDL2,QDL3}. Scalable ``on-chip" single emitter laser with ultra-low threshold power could have wide application in modern quantum technologies such as quantum information processing, quantum computing, quantum metrology, and analysis of complex quantum network \cite{QIQC1,QIQC2,QIQC3}. 

Semiconductor Quantum Dots (QDs) have emerged as a potential candidate for generating ``on-chip" quantum light sources. Owing to recent developments, different types of high finesse and low mode volume structures like toroidal, micropillar, and photonic crystal cavities have also been fabricated. The existing technology can also grow a QD inside the microcavity at a desired location \cite{QDF1,QDF2,QDF3,QDF4}. Various single emitter microlasers using QD embedded in a photonic crystal cavity \cite{QDL3,QDPCC2,QDPCC3,QDPCC4,QDPCC5} and QD coupled with coplanar microwave cavity have been realized recently \cite{QDL2,QDCPC2,QDCPC3}.
In solid-state semiconductor systems, lattice vibrations are inevitable due to the finite ambient temperature. Particularly, longitudinal acoustic phonon interactions with the QD exciton states significantly modify the system dynamics. The interactions between phonon modes and exciton lead to dephasing in coupled dynamics of exciton-photon interaction as well as off-resonant cavity mode feeding \cite{DPHZ1,DPHZ2,OFFCF1,OFFCF2,OFFCF3}. Henceforth, the microlaser dynamics are significantly modified. Phonon-assisted population inversion in QD using an incoherent and coherent pump has also been observed.

In the single-emitter single-photon laser, the lasing action occurs due to the stimulated emission of a single-photon in cavity mode. Similar to single-photon lasers, it has been predicted that the stimulated two-photon emission could also be used for two-photon lasing when single-photon emission is negligible in gain medium \cite{TWOPH}. The two-photon laser has been demonstrated by D. J. Gauthier {\it et al.} \cite{TWOPL} using strongly driven two-level atoms as the gain medium and probing by a weak field having frequency resonant to one of the sidebands. Recently the degenerate two-photon lasing in a single QD using an effective Hamiltonian approach has been proposed \cite{INTPL1,INTPL2}.  The experimental realization of degenerate two-photon maser has been achieved using Rydberg atoms passing through cavity \cite{TPM}. Further, the theory of a nondegenerate two-mode two-photon laser has also been proposed for inverted atoms passing through the cavity \cite{OTTML}. It has been predicted that in such a laser, one mode can play a role in enhancing and reducing gain in the second mode. Further, such laser can show entanglement between a large number of photons in two modes \cite{TMENT}. We also notice there has been some experimental success in the construction of a two-mode microlaser \cite{EXTML}, where two orthogonal polarized modes in a micropillar cavity are coupled with exciton states having angular momenta $\pm1$. A strong correlation between two modes and superthermal photon bunching has been predicted in such two-mode laser \cite{BANCH}. Though some success has been achieved for two-photon and two-mode lasers, the nondegenerate two-photon laser has not yet been investigated in the QD-cavity system. In this paper, we propose a scheme for a nondegenerate two-photon laser using a single QD embedded in a two-mode microcavity. 
 
 The paper is organized as follows. In Sec. \ref{sec:model}, we introduce the QD-cavity model system along with the polaron master equation for both coherent and incoherent pumping mechanisms. In Sec. \ref{sec:incoh} and \ref{sec:coh}, we analyze the two-mode laser in terms of QD population, cavity field parameters, and environment temperature for incoherent and coherent pumping. By choosing the suitable parameter regime, Sec. \ref{sec:entangled} displays the continuous two-mode entanglement in our system. Finally, Sec. \ref{conclud} is devoted to conclusions.                                                  

\section{MODEL SYSTEM}
\label{sec:model}
In this work, we investigate the feasibility of two-mode two-photon microlaser and entanglement generation in the semiconductor cavity quantum electrodynamics system. We consider a system composed of a single QD embedded in a two-mode high-quality photonic microcavity. In the neutral QD, the bound state of the electron-hole pair produces discrete energy levels. The relevant energy levels of the QD consists of a biexciton state $|u\rangle$, two single exciton states $|x\rangle$, $|y\rangle$, and a ground state $|g\rangle$, respectively, as  shown in Fig.\ref{Fig.1}. The frequencies of the excited states  $|u\rangle,|x\rangle$ and $|y\rangle$ are denoted by $\omega_{u}$, $\omega_{x}$ and $\omega_{y}$, respectively. The energy difference between two exciton states is denoted by fine-structure splitting $\delta_{x}$= $\omega_{x}-\omega_{y}$ and biexciton binding energy is defined by $\Delta_{xx}$= $\omega_{x}+\omega_{y}-\omega_{u}$. The QD energy levels form a diamond-like structure due to optically forbidden transition between two bright exciton states.
\begin{figure}[h]
   \includegraphics[scale=0.45]{fig_1}
    \caption{\label{fig:epsart}The energy level diagram for QD-cavity model. Left blue arrows indicating incoherent (coherent) pumping from $|g\rangle \rightarrow |u\rangle$ with pumping rates $\eta_{1}~(\Omega_1)$ and $\eta_{2}~(\Omega_2)$. Right green and red arrows indicating cavity coupling with $|y\rangle \rightarrow |g\rangle$ and $|u\rangle \rightarrow |y\rangle$ with coupling constant $g_{1}$ and $g_{2}$ and  frequencies $\omega_{1}$ and $\omega_{2}$. Here $\Delta_{xx}$ and $\delta_{x}$ stand for the biexciton binding energy and fine structure splitting energy between two exciton states. The $\Delta_{1}$ and $\Delta_{2}$ are the effective detunings for the first and second cavity modes.}
    \label{Fig.1}
\end{figure}
 It is clear from Fig.\ref{Fig.1} that two transition pathways exist between $|u\rangle$ and $|g\rangle$: one path is described by $|u\rangle \leftrightarrow |x\rangle \leftrightarrow |g\rangle$ and another represented by $|u\rangle \leftrightarrow |y\rangle \leftrightarrow |g\rangle$. We choose the quantum path $|u\rangle \leftrightarrow |x\rangle \leftrightarrow |g\rangle$, using H-polarized fields, to produce population inversion in the system. Here we consider incoherent pumping for $|g\rangle \rightarrow |x\rangle$ and $|x\rangle \rightarrow |u\rangle$ transition with pumping rate $\eta_{1}$ and $\eta _{2}$, respectively. The cavity has two V-polarized non-degenerate modes with frequencies $\omega_{1}$ and $\omega_{2}$. We consider transitions $|y\rangle\rightarrow |g\rangle$ and $|u\rangle \rightarrow |y\rangle$ are coupled with cavity modes with coupling constant $g_{1}$ and $g_{2}$.  The biexciton state may decay to ground state through one of the two exciton states by the emission of a horizontally (H) or vertically (V) polarized photons \cite{XYPOL}. 
\subsection{Master equation for Incoherent Pumping}
The Hamiltonian for the dot-cavity system in the rotating frame with frequency $\omega_y$ is written as
\begin{eqnarray}
H &&= \hbar\delta_{x}\sigma_{xx}- \hbar(\Delta_{xx}-\delta_{x})\sigma_{uu}\nonumber\\ 
&&- \hbar\delta_{1}a_{1}^{\dagger}a_{1} - \hbar(\Delta_{xx}-\delta_{x}+\delta_{2})a_{2}^{\dagger}a_{2} \nonumber\\
 &&+\hbar g_{1}(\sigma_{yg} a_{1}+ \sigma_{gy} a_{1}^{\dagger})+\hbar g_{2}(\sigma_{uy} a_{2} + \sigma_{yu} a_{2}^{\dagger})\label{uhu},
\end{eqnarray}
where $\delta_{1} = \omega_{y} - \omega_{1}$,
$\delta_{2} = \omega_{u}-\omega_{y}-\omega_{2}\nonumber $ are  cavity detunings, and $a_1$, $a_2$ are field operators of the first and second mode, respectively. The QD projection operators are denoted by $\sigma_{ij}=\vert i\rangle \langle j\vert$.
In the semiconductor QD-cavity system, the interaction of the electrons with the lattice modes of vibration, \textit{i.e.}, the longitudinal acoustic phonons, plays an important role. Various phenomena such as  strong dephasing, off-resonant cavity feeding, and  the appearance of new features in Mollow triplets have been observed due to the phonon interaction \cite{POLARON1,POLARON2,POLARON3}.
The phonon interaction in QD-cavity dynamics can be incorporated into the system by including the following terms
\begin{equation}
H_{ph}=\hbar\sum_k\omega_kb_k^{\dag}b_k+\sum_{i=x,y,u}\lambda_{ik}\sigma_{ii}(b_k+b_k^{\dag}),
\end{equation}
where $\lambda_{ik}$ are exciton phonon coupling constants and $b_k$($b_k^{\dag}$) are annihilation(creation) operators for $k^{\text{th}}$ phonon mode of frequency $\omega_k$. The second term in $H_{ph}$ represents the interaction between excitons with the phonon reservoir. The total Hamiltonian of the QD-cavity system including exciton-phonon interaction becomes 
\begin{equation}
H_{T}= H + H_{ph}.
\end{equation}
However, it should be borne in mind that including all orders of exciton phonon interaction in the master equation of QD-cavity system has been very challenging task.
Although many perturbative approaches have been considered in the literature by truncating higher-order phonon interaction terms \cite{WPHON1,WPHON2,POLARON1}. The most successful master equation for the QD-cavity system utilize the polaron transformation where all orders of phonon interaction are taken into account \cite{POLARON1,POLARON2,POLARON3}. The Hamiltonian $H_{T}$ is transformed into $H^{'}$ using polaron transformation $H^{\prime}=e^PH_{T}\hspace{0.5mm} e^{-P}$ where
\begin{equation}
P=\sum_{i=x,y,u}\frac{\lambda_{ik}}{\omega_k}\sigma_{ii}\left(b_k^{\dag}-b_k\right).\nonumber
\end{equation}
The transformed Hamiltonian $H^{\prime}$ contains system, bath, and their interaction Hamiltonian as $H^{\prime}=H_s+H_b+H_{sb}$, where 
\begin{eqnarray}
H_{s} &&= \hbar\delta_{x}\sigma_{xx}- \hbar(\Delta_{xx}-\delta_{x})\sigma_{uu} - \hbar\Delta_{1}a_{1}^{\dagger}a_{1}\nonumber\\ 
&&- \hbar(\Delta_{xx}-\delta_{x}+\Delta_{2})a_{2}^{\dagger}a_{2}+\langle B\rangle X_{g},\\
H_b &&=\hbar\sum_k\omega_k b_k^{\dag}b_k,\\
H_{sb} &&=\xi_gX_g+\xi_uX_u.
\end{eqnarray}
Here, we redefine the cavity detunings into effective detunings $\Delta_1$ and $\Delta_2$ after including the polaron shifts $\sum_k\lambda_{ik}^2/\omega_k$. 
The system operators are defined by
\begin{eqnarray}
X_{g}&&=\hbar\left( g_{1}\sigma_{yg} a_{1} + g_{2}\sigma_{uy} a_{2} \right) + H.c.\\
X_{u}&&=i\hbar\left(g_{1}\sigma_{yg} a_{1} + g_{2}\sigma_{uy} a_{2}\right) + H.c. ,
\end{eqnarray}
and the phonon bath fluctuation operators are
\begin{eqnarray}
 \xi_{g} &&= \frac{1}{2}\left( B_{+} + B_{-} -2\langle B\rangle  \right)\\
 \xi_{u} &&= \frac{1}{2i}\left( B_{+} - B_{-} \right),
 \end{eqnarray}
where $B_{+}$, $B_{-}$ are the phonon  displacement operator. 
The phonon displacement operators are defined by 
\begin{equation}
B_{\pm} =  \exp\left[{\pm \sum_{k} \frac{\lambda_{k}}{\omega_{k}}\left( b_{k} - b_{k}^{\dagger}\right)}\right].\nonumber 
\end{equation}
The phonon displacement operator averaged over all closely spaced phonon modes at a temperature T obeys the relation $\langle B_{+}\rangle =\langle B_{-}\rangle = \langle B\rangle$ where
\begin{equation}
\langle B\rangle= \text{exp}\left[-\frac{1}{2}\int_0^{\infty}d\omega\frac{J(\omega)}{\omega^2}
\coth\left(\frac{\hbar\omega}{2K_bT}\right)\right].
\end{equation}
The phonon spectral density function $J(\omega)$ is given by $J(\omega)= \alpha_{p}\omega^3\exp[-\omega^2/2\omega_b^2]$ \cite{SPECT}, where the parameters $\alpha_p$ and $\omega_b$ are
the electron-phonon coupling and cutoff frequency, respectively. We have chosen model parameters  $\alpha_p=1.42\times10^{-3}/g_1^2$ and $\omega_b=10g_1$, which gives us the value of $\langle B\rangle=1.0$, $0.90$, and $0.73$ for $T=0K$, $5K$, and $20K$, respectively \cite{PARA1,PARA2}. 

Now, applying second order Born-Markov approximation, the master equation for reduced density matrix  for QD-cavity system $\rho_s$  can be written as \cite{POLARON1,POLARON2,POLARON3}
\begin{eqnarray}
\dot{\rho_s} &&=-\frac{i}{\hbar}[H_s,\rho_s]-{\cal L}_{ph}\rho_s-\frac{\kappa_{1}}{2}{\cal L}[a_{1}]\rho_s\nonumber
-\frac{\kappa_{2}}{2}{\cal L}[a_{2}]\rho_s\nonumber\\
&&-\sum_{i=x,y}\left(\frac{\gamma_1}{2}{\cal L}[\sigma_{gi}]
+\frac{\gamma_2}{2}{\cal L}[\sigma_{iu}]\right)\rho_s
 -\sum_{i=x,y,u}\frac{\gamma_d}{2}{\cal L}[\sigma_{ii}]\rho_s\nonumber\\
&&-\left(\frac{\eta_1}{2}{\cal L}[\sigma_{xg}]
+\frac{\eta_2}{2}{\cal L}[\sigma_{ux}]\right)\rho_s\label{meq},
\end{eqnarray}
where we phenomenologically incorporate the photon leakage from the cavity modes with rates $\kappa_{1}$, $\kappa_{2}$, and  spontaneous decay of exciton and biexciton states with rates $\gamma_1$, $\gamma_2$. 
In addition, we also include the pure-dephasing process with rate $\gamma_d$ in the master equation (\ref{meq}). This additional dephasing term incorporates the broadening of the zero-phonon line (ZPL) in QDs \cite{ZPL1,ZPL2}. The last term in master equation (\ref{meq}) is responsible for the incoherent pumping process with pumping rates $\eta_{1}$ and $\eta_{2}$.  In the master equation, $\cal L$  stands for Lindblad superoperator defined as 
${\cal L}[{\cal O}]\rho_s ={\cal O}^{\dagger}{\cal O} \rho_s - 2 {\cal O} \rho_s {\cal O}^{\dagger} +\rho_s \cal O^{\dagger}\cal O $. The term ${\cal L}_{ph}\rho_s$ represents the effect of phonon bath on the system dynamics. The explicit form of ${\cal L}_{ph}\rho_s$ in terms of previously defined  system operators is written as
\begin{equation}
{\cal L}_{ph}\rho_s=\frac{1}{\hbar^2}\int_0^{\infty}d\tau\sum_{j=g,u}G_j(\tau)[X_j(t),X_j(t,\tau)\rho_s(t)]+H.c.
\label{lph}
\end{equation}
where $X_j(t,\tau)=e^{-iH_s\tau/\hbar}X_j(t)e^{iH_s\tau/\hbar}$, with the polaron Green functions $G_g(\tau)=\langle B\rangle^2\{\cosh[\phi(\tau)]-1\}$ and $G_u(\tau)=\langle B\rangle^2\sinh[\phi(\tau)]$. 
The phonon Green's functions depend on phonon correlation function 
\begin{equation}
\phi(\tau)=\int_0^{\infty}d\omega\frac{J(\omega)}{\omega^2}
\left[\coth\left(\frac{\hbar\omega}{2K_bT}\right)\cos(\omega\tau)-i\sin(\omega\tau)\right],
\end{equation}
where $K_b$ and $T$ are Boltzmann constant and the temperature of phonon bath, respectively. We integrate master equation (\ref{meq}) numerically using Quantum Optics Toolbox \cite{QOTOOL}. 
\subsection{Master equation for Coherent Pumping}
In this subsection, we derive master equation for QD-cavity system considering coherent field for pumping. We consider a horizontally polarised coherent field having frequency $\omega_{p}$  to the $\vert g\rangle \leftrightarrow \vert x \rangle$ and $\vert x\rangle \leftrightarrow \vert u \rangle$ transition with Rabi frequency $\Omega_{1}$ and $\Omega_{2}$, respectively. The Hamiltonian of this system in a rotating frame with frequency $\omega_{p}$ can be written as
\begin{eqnarray}
H &&= \hbar\Delta_{p}\sigma_{xx}+\hbar(2\Delta_{p}-\delta_{x}-\Delta_{xx})\sigma_{uu}+\hbar(\Delta_{p}-\delta_{x})\sigma_{yy} \nonumber\\ &&+\hbar(\Delta_{p}-\delta_{x}-\Delta_{1})a_{1}^{\dagger}a_{1}\nonumber +\hbar(\Delta_{p}-\Delta_{xx}-\Delta_{2})a_{2}^{\dagger}a_{2} \nonumber\\
&&+\hbar \Omega_{1}(\sigma_{xg} + \sigma_{gx})+\hbar \Omega_{2}(\sigma_{ux}+ \sigma_{xu}) + \hbar g_{1}(\sigma_{yg} a_{1}\nonumber\\
&&+ \sigma_{gy} a_{1}^{\dagger})+\hbar g_{2}(\sigma_{uy} a_{2} + \sigma_{yu} a_{2}^{\dagger}) + H_{ph},
\label{hhcoh}
\end{eqnarray}
where $\Delta_{p}= \omega_{x} - \omega_{p}$ and all others parameters are same as previous definitions.
As mentioned earlier, we make the previously defined polaron transformation.
After using polaron transformation, we get the system Hamiltonian and other operators as follows:
\begin{eqnarray}
H_{s} &&= \hbar\Delta_{p}\sigma_{xx}+\hbar(2\Delta_{p}-\delta_{x}-\Delta_{xx})\sigma_{uu}+\hbar(\Delta_{p}-\delta_{x})\sigma_{yy}    \nonumber\\ &&+\hbar(\Delta_{p}-\delta_{x}-\Delta_{1})a_{1}^{\dagger}a_{1} +\hbar(\Delta_{p}-\Delta_{xx}-\Delta_{2})a_{2}^{\dagger}a_{2}\nonumber\\
&&+\langle B\rangle X_{g},\label{hcoh}\\
X_{g}&&=\hbar\left(\Omega_{1}\sigma_{xg}+\Omega_{2}\sigma_{ux}+g_{1}\sigma_{yg} a_{1} + g_{2}\sigma_{uy} a_{2}\right) + H.c\label{xgcoh}\\
X_{u}&&=i\hbar\left(\Omega_{1}\sigma_{xg}+\Omega_{2}\sigma_{ux}+g_{1}\sigma_{yg} a_{1} + g_{2}\sigma_{uy} a_{2}\right)+H.c.\label{xucoh}
\end{eqnarray}
We use the defined system operators in the master equation (\ref{meq}) for coherently pumped QD by neglecting the incoherent pumping terms present in the equation.

\section{TWO-MODE LASING USING INCOHERENT PUMPING}
\label{sec:incoh}
\begin{figure}[h]
   \includegraphics[scale=0.34]{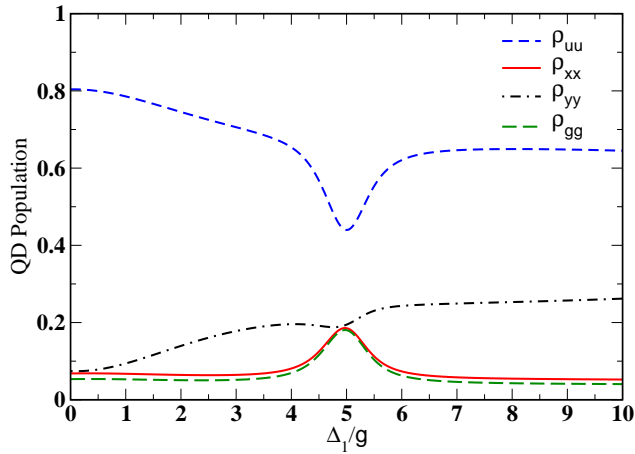}
    \caption{The steady state populations in QD states $|u\rangle$(blue short dash),$|x\rangle$(red solid),$|y\rangle$(black short dash dot), $|g\rangle$(green long dash) for phonon bath temperature T=5K. Other parameters are cavity leakage $\kappa_{1}=\kappa_{2}=0.1g$, cavity field couplings $g_{1}=g_{2}=g$, spontaneous decay rates $\gamma_{1}=\gamma_{2}=0.01g$, pure dephasing rate $\gamma_{d}=0.01g$, biexciton binding energy $\Delta_{xx}=15g$, hyperfine energy gap between excitons $\delta_{x}=-g$, detuning of second cavity mode $\Delta_{2}= -5g$ and $\eta_{1}=\eta_{2}=0.5g$.}
    \label{Fig.2}
\end{figure}

\begin{figure}[h]
   \includegraphics[scale=0.34]{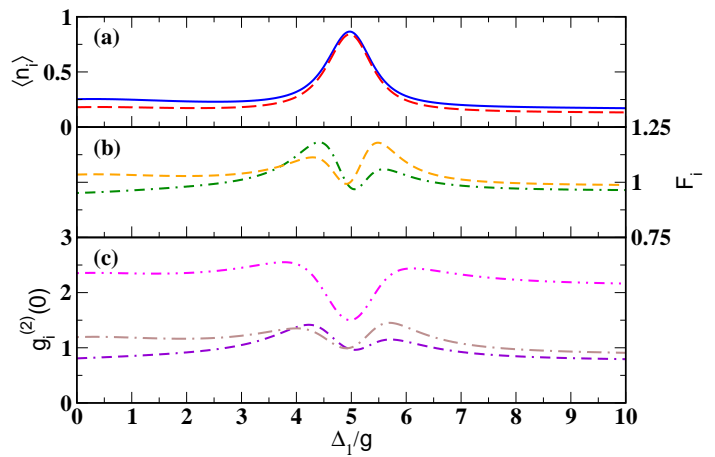}
    \caption{ a) The average photon number of first and second mode $\langle n_{1}\rangle$(blue solid), $\langle n_{2}\rangle $(red long dash), b) Fano factors $F_{1}$ (green short dash dot), $F_{2}$ (orange short dash), and c) Second order photon correlation function with zero time delay $g_{1}^{(2)}(0)$ (violet two short dash dot), $g_{2}^{(2)}(0)$ (brown long dash dot), $g_{12}^{(2)}(0)$(magenta short dash two dot) as a function of first cavity mode detuning $\Delta_{1}$ for temperature T=5K. All other parameters are same as Fig.$\ref{Fig.2}$.
    }
    \label{Fig.3}
\end{figure}

\begin{figure}[h]
   \includegraphics[scale=0.34]{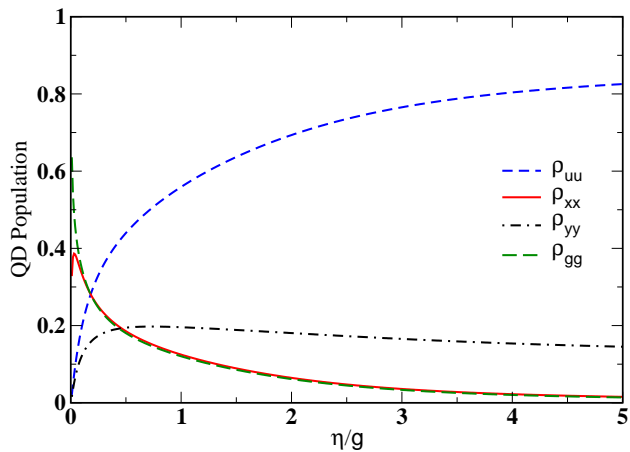}
    \caption{The steady state populations in QD states $|u\rangle$(blue short dash), $|x\rangle$(red solid), $|y\rangle$(black short dash dot), $|g\rangle$(green long dash) vs incoherent pumping rate $\eta$ for phonon bath temperature T=5K. All other parameters are same as Fig.$\ref{Fig.2}$ except $\Delta_{1}= 5g$ and $\eta_{1}=\eta_{2}= \eta$.}
    \label{Fig.4}
\end{figure}

\begin{figure}
   \includegraphics[scale=0.34]{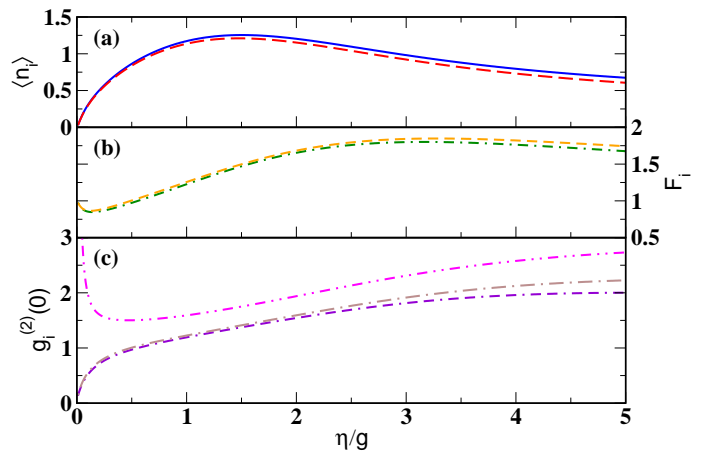}
    \caption{a) Average photon number for first and second mode $\langle n_{1}\rangle$(blue solid), $\langle n_{2}\rangle $(red long dash), b)Fano factors $F_{1}$(green short dash dot), $F_{2}$(orange short dash ), c) Second order correlations function with zero time delay $g_{1}^{(2)}(0)$(violet two short dash dot), $g_{2}^{(2)}(0)$(brown long dash dot), $g_{12}^{(2)}(0)$(magenta short dash two dot) as a function of $\eta$ for temperature T=5K. All other parameters are same as Fig.($\ref{Fig.2}$) except $\Delta_{1}= 5g$ and $\eta_{1}=\eta_{2}= \eta$.
    }
    \label{Fig.5}
\end{figure}

For incoherently pumped QD, we evolve the master equation (\ref{meq}) numerically and then calculate various QD populations and cavity field parameters in the steady-state condition. The results are shown in Fig.\ref{Fig.2}-\ref{Fig.5} for phonon bath temperature T=5K with other experimentally compatible parameters. We consider biexciton binding energy $\Delta_{xx}=15g_{1}$ and fine structure splitting of excitons $\delta_{x}=-g_{1}$. 
We explore the feasibility of achieving population inversion in the biexciton state by applying continuous incoherent pumping. Such incoherent pumping has been experimentally achieved through continuous excitation of wetting layer \cite{PUMP1} and has been widely implemented in single QD lasers.

In Figure $\ref{Fig.2}$, we demonstrate the variation of steady state populations in various QD states  with the detuning $\Delta_1$ from first cavity mode.
We find the population in  $|u\rangle$ state is larger than populations in other exciton states. This high population in biexciton state is due to the large detuning, $\vert\Delta_{2}\vert\gg g_2$, which prevent population transfer from $|u\rangle$ to $|y\rangle$. This situation is similar to the population inversion in the regular lasing action.
The efficient two-photon lasing can be attained through two-photon resonant transition from biexciton to the ground state.
We notice from Fig.$\ref{Fig.2}$ that at two-photon resonant condition $\Delta_1=-\Delta_2$, the population in $\vert u\rangle$ gets dip and the population in $\vert g\rangle$ gets a peak due to maximum transfer population $\vert u\rangle$ to $\vert g\rangle$.
Generally, the two-photon process is a weak nonlinear process; this can be enhanced by suppressing the single-photon process. The single-photon transitions are minimised by considering large exciton-field detunings $\vert\Delta_{2}\vert\gg g_2$.
Therefore even at single photon resonance condition $\Delta_{1}=0$  for $|y\rangle \rightarrow |g\rangle$ transition, single photon emission is  not significant because of very less population in $|y\rangle$ state.
Under above conditions, the probabilities of single-photon emission become very low. The two-photon resonance condition $\Delta_{1} + \Delta_{2}=0$ can be simplified as $ \omega_{u}=\omega_{1}+\omega_{2}$, therefore cavity induced decay of biexciton $|u\rangle$ to ground state $|g\rangle$ generates a pair of  non degenerate photons of frequencies $\omega_{1}$ and $\omega_{2}$. We notice that the $|y\rangle$ state population increases monotonically with a dip at two-photon resonance on increasing $\Delta_{1}$. In single-photon resonance condition, population in $|y\rangle$ state can easily transfer to the $|g\rangle$ state which explain the lowest population in $|y\rangle$ state at $\Delta_{1}=0$. On increasing $\Delta_{1}$, the transition $|y\rangle \rightarrow |g\rangle $ becomes off-resonant which leads to increasing population in $|y\rangle$ state.
Fig.$\ref{Fig.2}$ also exhibit the  identical variation of populations in $|x\rangle$ and $|g\rangle$ states.

The Fig.$\ref{Fig.3}$ depicts the statistics of the generated photons as a function of the first cavity mode detuning. 
In Fig.$\ref{Fig.3}$(a), we show the average photon number $\langle n_{1}\rangle=\langle a_1^{\dag}a_1\rangle$ and $\langle n_{2}\rangle=\langle a_2^{\dag}a_2\rangle$. 
The average numbers of photons in both modes remain same having a peak at the two-photon resonance.
The identical nature of the curves can be explained as photons are generated in biexciton-exciton cascaded decay. Under two-photon resonance condition, no individual photon emission occurs, but a photon pair is emitted. Hence each photon goes to the corresponding cavity mode with an equal probability.
The average number of photon peaks is consistent with the sudden decay of the $|u\rangle$ population as shown in Fig.$\ref{Fig.2}$. We analyze the Fano factor of the cavity field by taking the ratio between variance of photon number and the average photon number in each cavity mode. In terms of the cavity mode operators, the Fano factor of the first and second cavity mode can be written as
\begin{eqnarray}
F_{1} &&= (\langle n_{1}^{2}\rangle-\langle n_{1}\rangle^{2})/\langle n_{1}\rangle \\
F_{2} &&= (\langle n_{2}^{2}\rangle-\langle n_{2}\rangle^{2})/ \langle n_{2}\rangle.
\end{eqnarray}
Fig.$\ref{Fig.3}$(b) shows that both $F_{1}$ and $F_{2}$ have a dip near two-photon resonance. Otherwise, both have a constant value near unity. This minimum value of Fano factors indicates the suppression of noise in the cavity at two-photon resonance conditions.
Next we analyze the generated optical field in terms of the second-order correlation function. We follow the definition of second-order photon correlation functions with zero time delay for the two-mode cavity \cite{G2} which can be defined as
\begin{eqnarray}
g_{1}^{(2)}(0) &&= \langle a_{1}^{\dagger2}a_{1}^{2}\rangle/\langle a_{1}^{\dagger}a_{1}\rangle^{2} \\
g_{2}^{(2)}(0) &&= \langle a_{2}^{\dagger2}a_{2}^{2}\rangle/\langle a_{2}^{\dagger}a_{2}\rangle^{2} \\
g_{12}^{(2)}(0) &&= \langle a_{1}^{\dagger} a_{2}^{\dagger}a_{2}a_1\rangle/\langle a_{1}^{\dagger}a_{1}\rangle \langle a_{2}^{\dagger}a_{2}\rangle.
\end{eqnarray}
Fig.$\ref{Fig.3}$(c) shows that all the second-order correlations have a minimum value at the two-photon resonance condition. At two-photon resonance, $g_i^{(2)}(0)$ is almost one which implies that in each mode generated field is almost coherent in nature. However, the cross-correlation $g_{12}^{(2)}(0)$ remains more than one, indicating super-Poissonian nature.

In Fig.$\ref{Fig.4}$, we demonstrate the variation of steady-state QD populations as a function of incoherent pumping rate. The $|u\rangle$ state population increases with $\eta$ and saturate for a very high pumping rate. We notice from Fig.$\ref{Fig.4}$ that populations in $|g\rangle$ state and $|x\rangle$ state remain same. Further, the populations are higher for low pumping rate, and then become negligible on increasing pumping rate.  The population in the $|y\rangle$ state remains almost constant after population inversion is achieved in system. Clearly, for a very small pumping rate the population inversion occurs in the system. Further, the population inversion grows and saturates for a high pumping rate.
In Fig.$\ref{Fig.5}$, we have plotted previously defined cavity parameters by changing the pumping rate. The average photon number in both cavity modes increases and acquires maximum on increasing pump rate and then decreases for further higher pumping rate, resulting from a single emitter self-quenching effect \cite{QUENCH1}. The Fano factor of the cavity modes has the values smaller than one before inversion is achieved after that the vaules are larger than one. 
The second-order correlations for cavity modes $g_{1}^{(2)}(0)$ and $g_{2}^{(2)}(0)$ grow with incoherent pumping rate, then values approach to two for high pumping rates as shown in Fig.($\ref{Fig.5}$)(c). The generated field exhibit sub-Poissonian   photon statistics ($g_{i}^{(2)}(0)<1$) for low pumping rate  and then  become Poissonian ($g_{i}^{(2)}(0) = 1$) for moderate pumping rate $0.5g<\eta<1.0g$. In the case of higher pumping rate cavity field shows thermal photon statistics ($g_{i}^{(2)}(0) = 2$). This indicate on increasing incoherent pump photon emission becomes uncorrelated.
The inter-modal correlation $g_{12}^{(2)}(0)$ has values larger than one. Further, for low pumping rate, $g_{12}^{(2)}(0)$ first decreases and attains a minimum value, when stimulated emission becomes maximum, then again increases with increasing pumping rate. This observation suggests that all the correlation gets destroyed by increasing the pumping rate after a specific limit.

Next, we derive laser rate equation using quantum laser theory developed by Scully and Lamb \cite{QTL1,LRE2}. For that we simplify ${\cal L}_{ph}\rho_s$ term in master equation under off resonant condition $\vert\Delta_1\vert\gg g_1$ and $\vert\Delta_2\vert\gg g_2$ \cite{EFME}. The simplified master equation takes the form,
\begin{align}
\dot{\rho_s} &=-\frac{i}{\hbar}[H_{eff},\rho_s]-\sum_{i=x,y}\left(\frac{\gamma_1}{2}{\cal L}[\sigma_{gi}]
+\frac{\gamma_2}{2}{\cal L}[\sigma_{iu}]\right)\rho_s\nonumber\\
&-\frac{\kappa_{1}}{2}{\cal L}[a_{1}]\rho_s -\frac{\kappa_{2}}{2}{\cal L}[a_{2}]\rho_s -\sum_{i=x,y,u}\frac{\gamma_d}{2}{\cal L}[\sigma_{ii}]\rho_s\nonumber\\
&-\left(\frac{\eta_1}{2}{\cal L}[\sigma_{xg}] +\frac{\eta_2}{2}{\cal L}[\sigma_{ux}]\right)\rho_s -\frac{1}{2}( \Gamma_{2}^{+}{\cal L}[\sigma_{yu}a_{2}^{\dagger}]\nonumber\\
&+\Gamma_{2}^{-}{\cal L}[\sigma_{uy}a_{2}] +\Gamma_{1}^{+}{\cal L}[\sigma_{gy}a_{1}^{\dagger}] + \Gamma_{1}^{-}{\cal L}[\sigma_{yg}a_{1}])\\
&-\frac{\Gamma_{ug}}{2}(\sigma_{ug}a_{1}a_{2}\rho_s -2\sigma_{yg}a_{1}\rho_s a_{2}\sigma_{uy} + \rho_s\sigma_{ug}a_{1}a_{2})\nonumber\\
&-\frac{\Gamma_{gu}}{2}(\sigma_{gu}a_{1}^{\dagger}a_{2}^{\dagger}\rho_s -2\sigma_{yu}a_{2}^{\dagger}\rho_s a_{1}^{\dagger}\sigma_{gy} + \rho_s\sigma_{gu}a_{1}^{\dagger}a_{2}^{\dagger})\nonumber
\label{inco_simple_ms}
\end{align}
where the effective Hamiltonian
\begin{align}
H_{eff} &= H_{s} + \hbar (\delta_{2}^{+}\sigma_{uu}a_{2}a_{2}^{\dagger} +\delta_{2}^{-}\sigma_{yy}a_{2}^{\dagger}a_{2} +\delta_{1}^{+}\sigma_{yy}a_{1}a_{1}^{\dagger}\nonumber\\ 
&+\delta_{1}^{-}\sigma_{gg}a_{1}^{\dagger}a_{1}) +\hbar\Omega_{12}(\sigma_{ug}a_{1}a_{2} + \sigma_{gu}a_{1}^{\dagger}a_{2}^{\dagger}),
\end{align}
and various additional detunings, effective coupling, phonon-induced rates
\begin{align}
\delta_{i}^{\pm} &= g_{i}^{2}\langle B\rangle^{2}Im\left[ \int_{0}^{\infty} d\tau (e^{\phi(\tau)}-1)e^{\pm i\Delta_{i}\tau}\right]\nonumber\\
\Gamma_{i}^{\pm} &= 2g_{i}^{2}\langle B\rangle^{2}Re\left[ \int_{0}^{\infty} d\tau (e^{\phi(\tau)}-1)e^{\pm i\Delta_{i}\tau}\right]\nonumber\\
\Omega_{12} &= -\frac{i}{2}g_{1}g_{2}\langle B\rangle^{2}\left[\beta_{1} - \beta_{2}^{*}\right]\nonumber\\
\Gamma_{ug} &= g_{1}g_{2}\langle B\rangle^{2}\left[\beta_{1} + \beta_{2}^{*}\right]\nonumber\\
\Gamma_{gu} &= g_{1}g_{2}\langle B\rangle^{2}\left[\beta_{1}^{*} + \beta_{2}\right]\nonumber\\
\beta_{1} &= \int_{0}^{\infty} d\tau (e^{-\phi(\tau)}-1)e^{- i\Delta_{1}\tau}\nonumber\\
\beta_{2} &= \int_{0}^{\infty} d\tau (e^{-\phi(\tau)}-1)e^{i\Delta_{2}\tau}.
\end{align}

\begin{figure}[h]
   \includegraphics[scale=0.35]{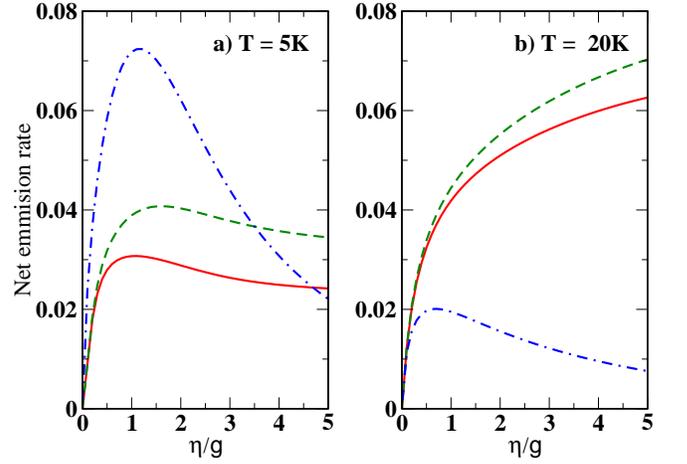}
    \caption{ First mode single-photon emission rate(red solid), Second mode single-photon emission rate(green dash) and two-photon stimulated emission rate(blue dash dot) using $\Delta_{1}= 5g_{1}$, $\eta_{1}=\eta_{2}=\eta$ for a)T=5K and b) T=20K with all other parameters same as Fig.$\ref{Fig.2}$.
   }
    \label{Fig.6}
\end{figure}
The two-photon laser rate equation is given by
\begin{align}
\dot{P}_{n,m} &=-\alpha_{n,m} P_{n,m}+G^{11}_{n-1,m-1}P_{n-1,m-1}  \nonumber\\
&+G^{10}_{n-1,m}P_{n-1,m}+G^{01}_{n,m-1}P_{n,m-1} \nonumber\\
&+A^{11}_{n+1,m+1}P_{n+1,m+1}+A^{10}_{n+1,m}P_{n+1,m}\nonumber\\
&+A^{01}_{n,m+1}P_{n,m+1}+ \kappa_{1}(n+1)P_{n+1,m}\nonumber\\& - \kappa_{1}nP_{n,m} + \kappa_{2}(m+1)P_{n,m+1} - \kappa_{2}mP_{n,m}, 
\label{inco_rate}
\end{align}
where $P_{n,m}=\sum_i\langle i,n,m|\rho_s|i,n,m\rangle$, and other terms on the right hand side $\alpha_{n,m}P_{n,m}=\sum_i\alpha_{i,n,m}\langle i,n,m|\rho_s|i,n,m\rangle$, $G^{\alpha\beta}_{n,m}P_{n,m}=\sum_iG^{\alpha\beta}_{i,n,m}\langle i,m,n|\rho_s|i,m,n\rangle$, $A^{\alpha\beta}_{m,n}P_{n,m}=\sum_iA^{\alpha\beta}_{i,n,m}\langle i,m,n|\rho_s|i,m,n\rangle$ with $i=g,x,y,u$ and $\alpha,\beta=0,1$. We mention that there could be some more terms in Eq.($\ref{inco_rate}$) corresponding to three photon and four photon emissions etc. However their contributions remain negligible under two-photon resonance. In Eq.(\ref{inco_rate}), the stimulated single-photon emission rates in first mode, second mode and two-photon emission rate when one photon is emitted in each mode, are given by  $\sum_{n,m} G_{n,m}^{10}P_{n,m}$, $\sum_{n,m} G_{n,m}^{(01)}P_{n,m}$ and $\sum_{n,m} G_{n,m}^{11}P_{n,m}$, respectively. Similarly,  single-photon absorption from first mode, second mode and two-photon absorption rates are given by  $\sum_{n,m} A_{n,m}^{10}P_{n,m}$, $\sum_{n,m} A_{n,m}^{01}P_{n,m}$ and $\sum_{n,m} A_{n,m}^{11}P_{n,m}$. We evaluate these rates numerically.  
We define the net single photon emission rate in the first mode by taking the difference between single photon emission rate and single photon absorption rate $\sum_{n,m}(G^{10}_{n,m}-A^{10})P_{n,m}$, the net single photon emission rate in second mode as $\sum_{n,m}(G^{01}_{n,m}-A^{01}_{n,m})P_{n,m}$. Similarly, the net emission rate of two-photon when one photon is emitted in each mode is defined as $\sum_{n,m} (G_{n,m}^{11}-A_{n,m}^{11})P_{n,m}$.

We plot net single-photon emission rates in first mode, second mode and net two-photon emission rate in Fig.\ref{Fig.6} at different temperatures. In Fig.$\ref{Fig.6}$(a), the stimulated two-photon emission rate dominates over single-photon emission rates in both modes and achieves maximum value for low pumping rate ($\eta\approx g$) at phonon bath temperature T=5K. 
However, on increasing incoherent pump the correlation in photon emission decreases as a result two-photon emission decreases and the single-photon emission in each mode start dominating. The single-photon emission rate of the second mode is higher than the first mode for a higher pumping rate. The higher two-photon emission rate at low pump indicates that the system acts like a two-photon gain medium. In Fig.$\ref{Fig.6}$(b), we plot net emission rates at higher phonon bath temperature T=20K.  The correlated two-photon emission decreases significantly and the independent single-photon emission rates in both modes increase.  This effect occurs due to the phonon-mediated process at higher temperatures. The two-photon process becomes negligible at a higher temperature, and the single-photon process gets enhanced due to phonon induced dephasing in the system. Comparing these two-photon and single-photon emissions, we conclude that the system show higher two-photon gain for low pump ($\eta\approx g$) at low temperatures.

\section{TWO-MODE two-photon LASING USING COHERENT PUMPING}
\label{sec:coh}
In this scheme, a coherent laser field resonantly excites the $|g\rangle \rightarrow |x\rangle$ transition. Therefore biexciton to exciton transition should be highly detuned due to the high biexciton binding energy. In this scenario, a very efficient way of biexciton generation is possible with the phonon-assisted transitions \cite{PAB1}. Previously, similar scheme has been demonstrated for high fidelity generation of biexciton state in QD \cite{BPWP}.
We solve the master equation $(\ref{meq})$ using Eqs.\ref{hcoh}, \ref{xgcoh}, \ref{xucoh} numerically to get the steady-state populations in different QD states and various cavity field parameters for coherent pumping.
\begin{figure}[h]
   \includegraphics[scale=0.32]{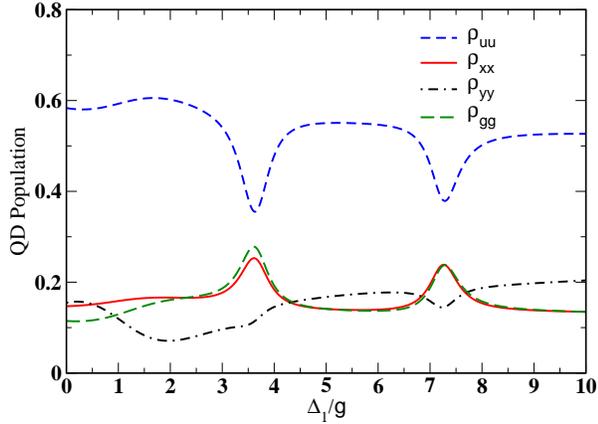}
    \caption{The steady state populations in the verious quantum dot energy states $|u\rangle$(blue short dash), $|x\rangle$(red solid), $|y\rangle$(black short dash dot), $|g\rangle$(green long dash) with first cavity mode detuning $\Delta_{1}$ for phonon bath temperature T=5K. Other parameters are same as Fig.($\ref{Fig.2}$) except $\Delta_{p} =0$ and $\Omega_{1} = \Omega_{2} = 2g$.}
    \label{Fig.7}
\end{figure}
In Fig.$\ref{Fig.7}$, we show the variation of the population in different QD states  with detuning $\Delta_{1}$ for coherent pumping. The population in biexciton state remains higher than in any other QD-states for typical values of system parameters, implying population inversion in the system.
We observe that the biexciton state population has two minima separated by $2\Omega_1$, when two-photon resonance condition is satisfied. This new feature can be explained using dressed state picture. On applying a resonant strong coherent field between the $\vert g\rangle$ and $\vert x\rangle$, the dressed states $\vert \pm\rangle = \frac{1}{\sqrt{2}}\left(\vert x\rangle\pm \vert g\rangle\right)$ are formed separated by $2\Omega_1$ in frequencies. Fig.$\ref{Fig.8}$ displays the equivalent dressed state picture of the system in the presence of strong coherent pumping.
\begin{figure}[h]
   \includegraphics[scale=0.45]{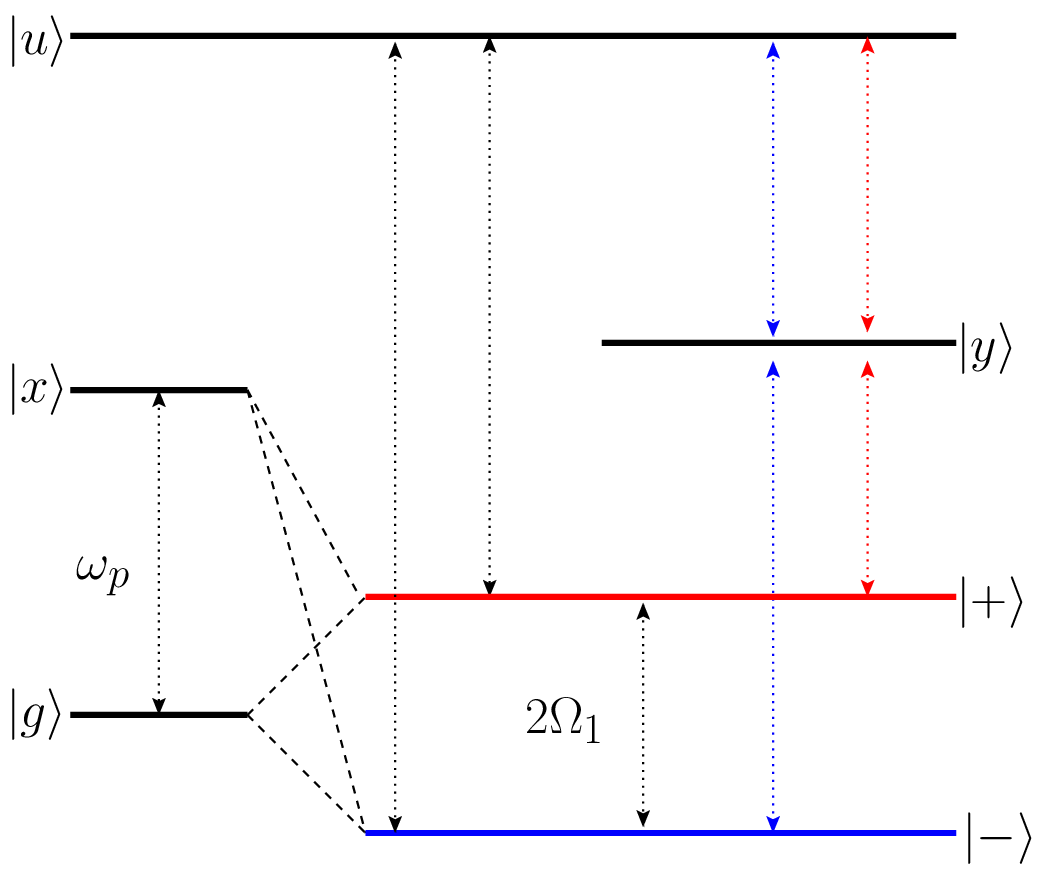}
    \caption{The strong optical field coupled with the $\vert g\rangle\leftrightarrow\vert x \rangle$ transition produces dressed states $\vert \pm\rangle$ with energy separation 2$\Omega_{1}$. Then these dressed states make a phonon-assisted pumping to the biexciton state denoted by black dotted lines. Now we can get two different values of $\Delta_{1}$ satisfying two-photon resonance condition denoted by red and blue dotted lines.}
    \label{Fig.8}
\end{figure} 
 The two dips in the population of biexciton state $\vert u\rangle$ correspond to the transitions between bare state $\vert u\rangle$ to dressed states  $\vert +\rangle$ and $\vert -\rangle$ through two-photon transitions in cavity modes, respectively. Therefore the two-photon emission occurs at two different values of the first mode detuning $\Delta_{1}=-\Delta_2\pm\Omega_{1}$.  The populations of $\vert g\rangle$ and $\vert x\rangle$ remain almost same and have maxima for these two values of $\Delta_1$ due to the population transfer from $\vert u\rangle\leftrightarrow\vert \pm\rangle$. The population in state $|y\rangle$ remains small with minimum at two-photon resonances.

\begin{figure}[h]
   \includegraphics[scale=0.32]{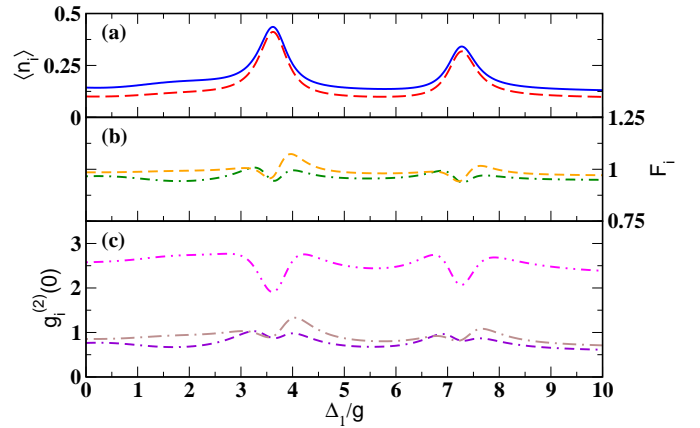}
    \caption{The variation of a) average photon number $\langle n_{1}\rangle$(blue solid), $\langle n_{2}\rangle $(red long dash), b) Fano factors $F_{1}$(green short dash dot), $F_{2}$(orange short dash), and c) Second order correlation $g_{1}^{(2)}(0)$(violet two short dash dot), $g_{2}^{(2)}(0)$(brown long dash dot), $g_{12}^{(2)}(0)$(magenta short dash two dot) with $\Delta_{1}$ for temperature T=5K. All other parameters are same as Fig.($\ref{Fig.7})$.}
    \label{Fig.9}
\end{figure}
The average photon numbers in the cavity modes have two prominent peaks with a separation of $2\Omega_1$ as shown in Fig.$\ref{Fig.9}$(a). The peaks in the average number of photons in cavity modes appear when two nondegenerate photons are emitted into the cavity modes by two-photon transition $\vert u\rangle\rightarrow |y\rangle\rightarrow\vert \pm\rangle$. We also notice that the two peaks in average photon number in each cavity mode are not identical. Although the separation is nearly $2\Omega_{1}$, the peaks in average photon number are not symmetrically placed around $\Delta_{1}=-\Delta_2$. Both peaks in the average number of photons are a little bit shifted to the higher frequency side due to the optical stark shifts in the QD energy levels \cite{STARK1,STARK2}. 
The values of Fano-factor for both cavity modes remain close to one and have minimum values when nondegenrate two-photon resonant  conditions are satisfied for $\Delta_{1}=-\Delta_2\pm\Omega_1$.
 The second order correlations, for each cavity mode $g_{i}^{(2)}(0)$ have values close to one. Hence the generated photons in each mode follow Poissonian statistics and indicate coherent field generation. Subsequently, intermodal correlation, $g_{12}^{(2)}(0)$, exhibits values larger than two except when two-photon resonant condition are satisfied where intermodal correlation has minimum value of $2$. Therefore, it is possible to construct two-mode two-photon lasing in single QD when two-photon resonant condition conditions are satisfied.
 
In the case of coherent pumping, we similarly found the effective master equation,
\begin{align}
\dot{\rho_s} &=-\frac{i}{\hbar}[H_{eff},\rho_s]-\sum_{i=x,y}\left(\frac{\gamma_1}{2}{\cal L}[\sigma_{gi}]
+\frac{\gamma_2}{2}{\cal L}[\sigma_{iu}]\right)\rho_s\nonumber\\
&-\frac{\kappa_{1}}{2}{\cal L}[a_{1}]\rho_s -\frac{\kappa_{2}}{2}{\cal L}[a_{2}]\rho_s -\sum_{i=x,y,u}\frac{\gamma_d}{2}{\cal L}[\sigma_{ii}]\rho_s\nonumber\\
&-\frac{1}{2}( \Gamma_{2}^{+}{\cal L}[\sigma_{yu}a_{2}^{\dagger}]+\Gamma_{2}^{-}{\cal L}[\sigma_{uy}a_{2}]\nonumber\\ 
&+\Gamma_{1}^{+}{\cal L}[\sigma_{gy}a_{1}^{\dagger}] + \Gamma_{1}^{-}{\cal L}[\sigma_{yg}a_{1}])\\
&-\frac{\Gamma_{ug}}{2}(\sigma_{ug}a_{1}a_{2}\rho_s -2\sigma_{yg}a_{1}\rho_s a_{2}\sigma_{uy} + \rho_s\sigma_{ug}a_{1}a_{2})\nonumber\\
&-\frac{\Gamma_{gu}}{2}(\sigma_{gu}a_{1}^{\dagger}a_{2}^{\dagger}\rho_s -2\sigma_{yu}a_{2}^{\dagger}\rho_s a_{1}^{\dagger}\sigma_{gy} + \rho_s\sigma_{gu}a_{1}^{\dagger}a_{2}^{\dagger})\nonumber\\
&-\frac{1}{2}( \Gamma_{2}^{p+}{\cal L}[\sigma_{xu}]+\Gamma_{2}^{p-}{\cal L}[\sigma_{ux}] +\Gamma_{1}^{p+}{\cal L}[\sigma_{gx}] + \Gamma_{1}^{p-}{\cal L}[\sigma_{xg}])\nonumber\\
&-\frac{\Gamma_{ug}^{p}}{2}(\sigma_{ug}\rho_s -2\sigma_{xg}\rho_s\sigma_{ux} + \rho_s\sigma_{ug})\nonumber\\
&-\frac{\Gamma_{gu}^{p}}{2}(\sigma_{gu}\rho_s -2\sigma_{xu}\rho_s \sigma_{gx} + \rho_s\sigma_{gu})\nonumber
\label{coh_simple_ms}
\end{align}
where the effective Hamiltonian with coherent pumping
\begin{align}
H_{eff} &= H_{s} + \hbar (\delta_{2}^{p+}\sigma_{uu} + (\delta_{2}^{p-} + \delta_{1}^{p+})\sigma_{xx} + \delta_{1}^{p-}\sigma_{gg}\nonumber\\
&+ \delta_{2}^{+}\sigma_{uu}a_{2}a_{2}^{\dagger} +\delta_{2}^{-}\sigma_{yy}a_{2}^{\dagger}a_{2} +\delta_{1}^{+}\sigma_{yy}a_{1}a_{1}^{\dagger}\nonumber\\ 
&+\delta_{1}^{-}\sigma_{gg}a_{1}^{\dagger}a_{1}) +\hbar\Omega_{12}(\sigma_{ug}a_{1}a_{2} + \sigma_{gu}a_{1}^{\dagger}a_{2}^{\dagger})\nonumber\\
&+\hbar\Omega^{p}(\sigma_{ug} + \sigma_{gu}),
\end{align}
and new symbols are defined as
\begin{align}
\delta_{2}^{p\pm} &= \Omega_{2}^{2}\langle B\rangle^{2}Im\left[ \int_{0}^{\infty} d\tau (e^{\phi(\tau)}-1)e^{\pm i\Delta_{p}^{'}\tau}\right]\nonumber\\
\delta_{1}^{p\pm} &= \Omega_{1}^{2}\langle B\rangle^{2}Im\left[ \int_{0}^{\infty} d\tau (e^{\phi(\tau)}-1)e^{\pm i\Delta_{p}\tau}\right]\nonumber\\
\Gamma_{2}^{p\pm} &= 2\Omega_{2}^{2}\langle B\rangle^{2}Re\left[ \int_{0}^{\infty} d\tau (e^{\phi(\tau)}-1)e^{\pm i\Delta_{p}^{'}\tau}\right]\nonumber\\
\Gamma_{1}^{p\pm} &= 2\Omega_{1}^{2}\langle B\rangle^{2}Re\left[ \int_{0}^{\infty} d\tau (e^{\phi(\tau)}-1)e^{\pm i\Delta_{p}\tau}\right]\nonumber\\
\Omega^{p} &= -\frac{i}{2}\omega_{1}\omega_{2}\langle B\rangle^{2}\left[\alpha_{1} - \alpha_{2}^{*}\right]\nonumber\\
\Gamma_{ug}^{p} &= \Omega_{1}\Omega_{2}\langle B\rangle^{2}\left[\alpha_{1} + \alpha_{2}^{*}\right]\\
\Gamma_{gu}^{p} &= \Omega_{1}\Omega_{2}\langle B\rangle^{2}\left[\alpha_{1}^{*} + \alpha_{2}\right]\nonumber\\
\alpha_{1} &= \int_{0}^{\infty} d\tau (e^{-\phi(\tau)}-1)e^{- i\Delta_{p}\tau}\nonumber\\
\alpha_{2} &= \int_{0}^{\infty} d\tau (e^{-\phi(\tau)}-1)e^{i\Delta_{p}^{'}\tau}, \Delta_{p}^{'} = \omega_{u} - \omega_{x} - \omega_{p}.\nonumber
\end{align}

\begin{figure}[h]
   \includegraphics[scale=0.35]{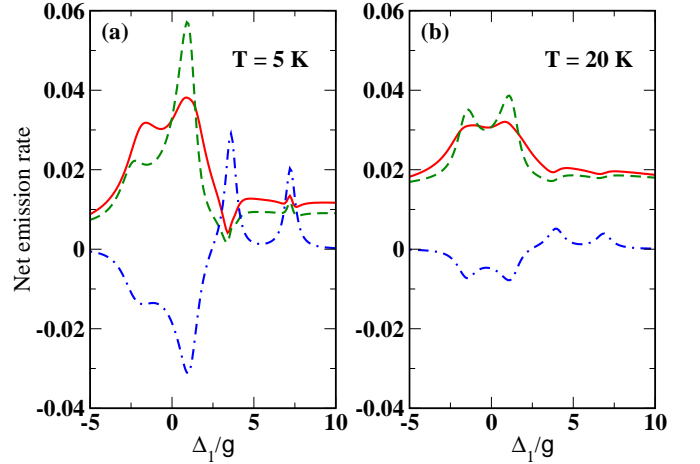}
    \caption{ First mode single-photon emission rate(red solid), Second mode single-photon emission rate(green dash), two-mode two-photon stimulated emission rate (blue dash dot) using $\Omega_{1}=\Omega_{2}= 2g_{1}$, for a)T=5K and b) T=20K with all other parameters same as Fig.$\ref{Fig.7}$.}\label{Fig.10}
\end{figure}
Following method discussed in previous section we derive laser rate equation (\ref{inco_rate}) for coherently pumped QD. Then we plot previously defined net emission rates as a function of first-mode detuning for two different temperatures in Fig.\ref{Fig.10}. 
From Fig.$\ref{Fig.10}$(a), we notice that the net stimulated single photon emission rate in the first and second modes show positive peaks when the single-photon transition $\vert y\rangle\rightarrow \vert \pm\rangle$ becomes resonant. These transition decrease population in $|y\rangle$ thus increasing population inversion between $|u\rangle$ and $|y\rangle$, therefore similar peaks are visible in net single-photon stimulated emission in second mode. On the other hand, the two-mode two-photon net emission rate displays negative dips around single-photon resonance, which indicates strong two-mode two-photon absorption. Therefore no appreciable change occurs in average number of photons in cavity modes. This scenario changes drastically around the two-photon resonance condition, where the two-mode two-photon net emission rate dominates over the single-photon emission rates. Under two-photon resonant condition $\Delta_1\approx-\Delta_2\pm\Omega_1$, the single-photon processes are highly detuned and net single-photon stimulated emission in both modes become very small, but the two-mode two-photon net stimulated emission has peaks. The double peaks come from the dressed states created by the coherent pumping. Further, for $\Delta_1\approx-\Delta_2-\Omega_1$, single-photon net stimulated emission rates becomes almost zero, therefore more suitable two-mode two-photon lasing conditions are achieved.
In Fig.$\ref{Fig.10}$(b), we plot net stimulated emission rates for higher temperature T = 20K. This figure depicts that the two-mode two-photon net emission rate gets suppressed, and the single-photon net emission rates get enhanced. This outcome is due to the domination of the phonon process at a higher temperature, which inhibits correlated two-photon emission in two-modes. In conclusion, the two-mode two-photon lasing for coherent pumping is possible at low temperatures.

\section{Generation of steady state two-mode entangled state}
\label{sec:entangled}
In this section, we consider two-photon resonant coherent pumping between $\vert g\rangle\rightarrow\vert x\rangle\rightarrow\vert u\rangle$ transition. Therefore, we select exciton pump detuning such that the two-photon resonant condition $\omega_u=2\omega_p$, i.e., $\Delta_{p}=\left(\Delta_{xx}+\delta_{x}\right)/2$ is satisfied.
\begin{figure}[h]
   \includegraphics[scale=0.32]{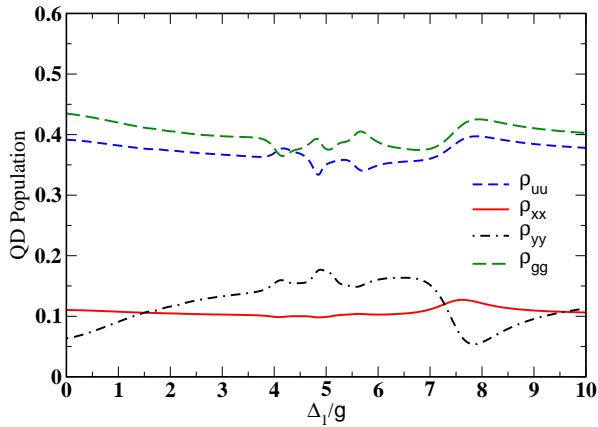}
    \caption{Shows the variation of steady state populations in quantum dot energy states $|u\rangle$(blue short dash), $|x\rangle$(red solid), $|y\rangle$(black short dash dot), $|g\rangle$(green long dash) as a function of $\Delta_{1}$ for same parameters as Fig.$\ref{Fig.7}$ except $\Delta_{p}=7g$.}
    \label{Fig.11}
\end{figure}

In Fig.$\ref{Fig.11}$, we show steady state populations in different energy states in QD. The populations in ground state $\vert g\rangle$ and biexciton state $|u\rangle$ remain almost equal and much larger than populations in $\vert x\rangle$ and $\vert y\rangle$. Further due to spontaneous decay of biexciton state the population in ground state $|g\rangle$ becomes larger than population in biexciton state $|u\rangle$. Thus there is no population inversion in the system. In the two-photon resonant pumping, the population in exciton state $\vert x\rangle$ remains small and the simultaneous absorption of two pump photons leads to $\vert g \rangle \leftrightarrow \vert u\rangle$ transition.  The biexciton state $|u\rangle$ decays to ground state $|g\rangle$ through cavity modes via two-mode two-photon resonant transition $|u\rangle\rightarrow|y\rangle\rightarrow|g\rangle$ when the two-photon resonance condition $\Delta_{1}+\Delta_{2} =0$ is satisfied.
\begin{figure}[h]
   \includegraphics[scale=0.32]{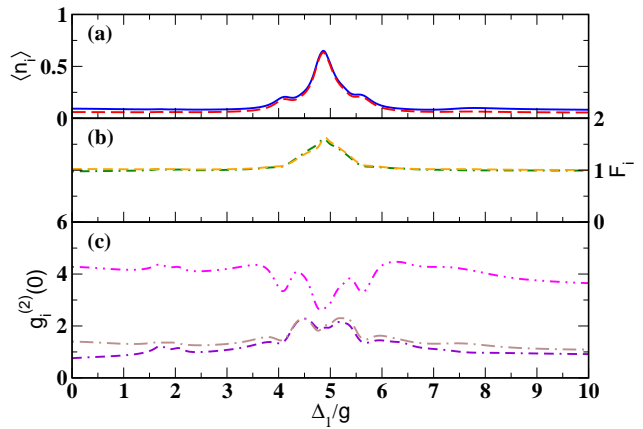}
    \caption{Display the variation of a) average photon number $\langle n_{1}\rangle$(blue solid), $\langle n_{2}\rangle $(red long dash) b) Fano factors $F_{1}$(green short dash dot), $F_{2}$(orange short dash ) c) second order corelations $g_{1}^{(2)}(0)$(violet two short dash dot), $g_{2}^{(2)}(0)$(brown long dash dot) , $g_{12}^{(2)}(0)$(magenta short dash two dot) as a function of $\Delta_{1}$ for same parameters as Fig.$\ref{Fig.9}$ except $\Delta_{p}=7g$.}
    \label{Fig.12}
\end{figure}
The average photon numbers in cavity modes have peaks at two-photon resonance conditions ( shown in Fig.$\ref{Fig.12}$(a)). The Fano-factors show a value greater than unity near the two-photon resonance, indicating that the fields generated in both modes have higher noise than coherent states. All the second-order correlations show similar behaviour as previous case except small oscillations due to coherence.

\begin{figure}
   \includegraphics[scale=0.35]{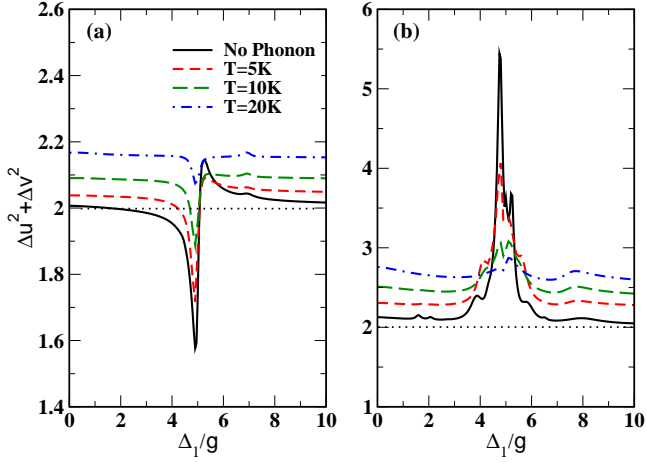}
    \caption{We have plotted sum of the variance of EPR like variable pair with first mode single photon detuning $\Delta_{1}$ for different bath temperatures. The Black dotted line indicates the margin for entanglement phenomenon according to the DGCZ criterion. All the parameters are the same as Fig.($\ref{Fig.11}$) except $\phi_{1} = \phi_{2} = -0.5$ with coherent pumping Rabi frequencies a)$\Omega_{1}=\Omega_{2}=0.5g$ and b)$\Omega_{1}=\Omega_{2}=2g$.
   }
    \label{Fig.13}
\end{figure}

Next, we discuss the steady state entanglement between the two non-degenerate cavity modes. We use the Duan-Giedke-Cirac-Zoller (DGCZ) criterion \cite{DGCZ1,DGCZ2} for the continuous variable entanglement. We define two EPR like variables for the field in cavity modes such as
\begin{equation}
 u =x_{1}+x_{2},v=p_{1}-p_{2},\label{EPRvar}
\end{equation}
where $x_{j},p_{j}$ for jth cavity mode are written in terms of the cavity operators
\begin{align}
 x_{j} &=\frac{1}{\sqrt{2}}(a_{j}^{\dagger}e^{-i\phi_{j}}+a_{j}e^{i\phi_{j}}),\\
 p_{j} &=\frac{i}{\sqrt{2}}(a_{j}^{\dagger}e^{-i\phi_{j}}-a_{j}e^{i\phi_{j}}),j=\{1,2\}.
\end{align}
According to the DGCZ criterion, two quantized modes of our cavity radiation are entangled if the quantum fluctuations of the operators $u$ and $v$ satisfy the inequality
\begin{equation}
\Delta u^{2} + \Delta v^{2} = \langle (u-\langle u\rangle)^{2}+(v-\langle v\rangle)^{2} \rangle < 2.
\end{equation}
The left hand side of the inequality can be rewritten in terms of experimentally measurable quantities
\begin{align}
\Delta u^{2} + \Delta v^{2} &= 2( 1+\langle a_{1}^{\dagger}a_{1}\rangle+\langle a_{2}^{\dagger}a_{2}\rangle +\langle a_{1}^{\dagger}a_{2}^{\dagger}\rangle e^{-i(\phi_{1}+\phi_{2})}\nonumber\\
&+\langle a_{1}a_{2}\rangle e^{i(\phi_{1}+\phi_{2})} -\langle a_{1}^{\dagger}\rangle\langle a_{1}\rangle -\langle a_{2}^{\dagger}\rangle\langle a_{2}\rangle\nonumber\\
&-\langle a_{1}^{\dagger}\rangle\langle a_{2}^{\dagger}\rangle e^{-i(\phi_{1}+\phi_{2})} -\langle a_{1}\rangle\langle a_{2}\rangle e^{i(\phi_{1}+\phi_{2})} ).
\end{align}
The Fig.$\ref{Fig.13}$ depicts the sum of the variance of two EPR like variables $\Delta u^{2} + \Delta v^{2}$ with $\Delta_{1}$ for different temperatures using typical values of system parameters. We use pump strength smaller than cavity mode coupling ($\Omega_1=\Omega_2=0.5g$) in Fig.\ref{Fig.13}(a) and larger than cavity mode coupling ($\Omega_1=\Omega_2=2.0g$) in Fig.\ref{Fig.13}(b). According to the DGCZ criterion, two cavity modes become entangled if  $\Delta u^{2} + \Delta v^{2}<2$, the critical value is shown by the black dotted line. It is clear that for two-photon pump strength smaller than cavity coupling ($\Omega_1=\Omega_2<g$) and at low temperature ($T<10K$) two-modes are entangled.
\section{CONCLUSIONS}
\label{conclud}
In conclusion, we have predicted the feasibility of two-mode two-photon lasing in a single quantum dot embedded in the two-mode microcavity system. We have considered the effect of exciton-phonon interaction using the polaron master equation formalism. For incoherently pumped quantum dot stimulated two-mode two-photon emission rate is dominating at low pump strength and at temperature $T=5K$. For coherently pumped quantum dot one can achieve large two-mode two-photon stimulated emission with negligible single-photon stimulated emission, which provide better conditions for realization of nondegenrate two-photon lasing in single quantum dot. We have also predicted that using two-photon resonant pump steady state two-mode entangled states can be generated.
The two-mode two-photon laser scheme may have potential applications in scalable quantum photonic technology in the field of quantum computation, quantum information, and large quantum network. The two-mode continuous-variable entanglement scheme can offer an  advantage in some situations in quantum information science. 
%\bibliography{reference}% Produces the bibliography via BibTeX.
\bibliography{paper}
\end{document}